# Extension Decisions in Open Source Software Ecosystem

Elmira Onagh[a,*], Maleknaz Nayebi[a]

[a]York Univerity, Toronto, Canada

## ABSTRACT

GitHub Marketplace is expanding by approximately 41% annually, with new tools; however, many additions replicate existing functionality. We study this phenomenon in the platform's largest segment, Continuous Integration (CI), by linking 6,983 CI Actions to 3,869 providers and mining their version histories. Our graph model timestamps every functionality's debut, tracks its adoption, and clusters redundant tools. We find that approximately 65% of new CI Actions replicate existing capabilities, typically within six months, and that a small set of first-mover Actions accounts for most subsequent forks and extensions. These insights enable developers to choose the optimal moment to launch, target unmet functionality, and help maintainers eliminate redundant tools. We publish the complete graph and dataset to encourage longitudinal research on innovation and competition in software ecosystems, and to provide practitioners with a data-driven roadmap for identifying emerging trends and guiding product strategy.

## 1. Introduction

Software Ecosystems (SECOs) drive innovation through collaboration, competition, and reuse (1; 2; 3). GitHub Marketplace illustrates this dynamic: developers share and refine automation tools mainly in the form of GitHub Actions, cutting costs and accelerating releases via open-source contributions (4; 5; 6). Like mobile app stores such as Apple's App Store and Google Play Store (7; 8), the Marketplace is a two-sided platform linking providers and users (9; 10). However, the audiences differ. Mobile stores serve mainly non-technical consumers and deliver closed, compiled apps (11; 12), which boosts security but limits transparency and extensibility (13; 14). GitHub Marketplace targets developers; The transparency of its tools and source code enables inspection, remixing, and rapid improvement.

The GitHub Marketplace, with its publicly accessible source code and activity traces, offers a rich setting for studying tool evolution (11). Yet, many Actions in crowded domains such as Continuous Integration (CI) duplicate one another. Explaining why developers create new but functionally similar Actions instead of reusing existing ones is key to understanding innovation, feature drift, and competition (4; 15). Developers choose Actions for both functional and non-functional qualities and frequently adapt them through forks and pull requests (16). This open, collaborative cycle makes GitHub Marketplace a living laboratory for research on software automation and innovation.

### 1.1. What is a GitHub Action?

Originally a CI/CD aid, GitHub Actions has grown into a general automation backbone, overtaking services like Travis CI (17). The catalog expanded from 7,878 Actions in April 2021 to 22,194 in March 2024 (16). Each Action bundles a job graph, automating various activities in YAML format and must include an `action.yaml` descriptor in their root repository (15).

Developers can automate workflows in their repositories by creating custom GitHub Actions, reusing those from other projects, or selecting from the GitHub Marketplace, which offers a range of prebuilt solutions. These Actions can be customized via forks and improved collaboratively through pull requests, fostering open-source innovation. Figure 1 illustrates how a publisher releases an Action to the Marketplace and how users integrate it into their workflows. Integration requires manually specifying the Action's identifier in a workflow YAML file (15), guided by instructions in the corresponding `action.yaml` file.

### 1.2. Why Study the GitHub Marketplace?

GitHub Marketplace exemplifies a SECO where diverse automation tools interact (4), making it a valuable case for studying system co-evolution (18; 19). Its focus on automation (e.g., Actions) highlights how innovation streamlines development, reduces costs, and enhances productivity (20; 21). The marketplace's open and dynamic nature directly relates to the broader challenges of Systems-of-Systems (SoS) and SECO engineering, positioning it as an ideal case study for analyzing SECO evolution from a Software Engineering perspective. GitHub Marketplace serves as a microcosm of software development, showcasing how new tools and features emerge, evolve, and compete.

Unlike proprietary platforms, GitHub offers access to source code, usage statistics, and contribution history, which we refer to as *Rich Open Data*, enabling empirical analysis of innovation patterns. This transparency allows researchers to track the life cycle of automation tools, study adoption trends, and examine the mechanisms behind feature innovation (22; 23). The Marketplace also exhibits *Extensive Redundancy* and a *Competitive Landscape*, where both tool providers and users often possess similar technical expertise. This contrasts with other software distribution platforms, where consumers typically lack the ability to assess or modify software. Full pipeline visibility, including access to source code, creates opportunities for component-based analysis and software reuse, potentially improving

---

eonagh@yorku.ca (E. Onagh); nayebi@yorku.ca (M. Nayebi)
ORCID(s):

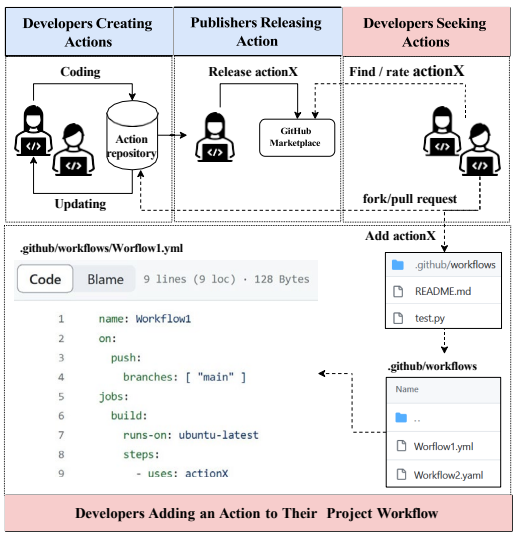

**Figure 1:** Actions are developed and shared on GitHub Marketplace, where they are integrated by others to automate workflows.

efficiency and reducing duplication. The prevalence of overlapping tools raises critical questions about competition, redundancy, and differentiation in open source ecosystems (4). Studying the GitHub Marketplace thus enables a deeper understanding of SECO evolution, feature emergence, and the dynamics of competition and collaboration. These insights position the Marketplace as a key platform for bridging empirical software engineering with real-world development practices.

To achieve the objectives of this study, we analyze the functionalities of GitHub Actions and their relationships in terms of shared capabilities (i.e., functional relationships). Additionally, we examine how these relationships have evolved over time, a process we term functional evolution, and identify the competitive factors driving this transformation within the SECO landscape. This approach sheds light on how automation tools and the GitHub Marketplace contribute to innovation in software development. Although our focus is on GitHub Actions, the patterns of functional evolution uncovered offer broader insights for developers and stakeholders across digital marketplaces. Our findings support improved developer productivity by offering actionable guidance for designing, refining, and maintaining tools in competitive and complementary environments. Beyond GitHub, the study provides implications for developers, platform managers, and researchers aiming to foster innovation, reduce redundancy, and drive sustainable growth in other SECOs. To this end, we address three core research questions (RQs) in the next section.

## 1.3. Research Questions

We focused on three research questions to evaluate the dynamics of the GitHub marketplace at the functional level:

**RQ1: How do the functional relationships among offerings in the GitHub Marketplace evolve within this SECO?**

To address this question, we analyzed the YAML files of Actions at two time snapshots, $t_0$ and $t_1$, to extract and categorize their functionalities. For each point in time, Actions were classified as either independent (having no functional overlap) or overlapping with other Actions. For overlapping Actions, we applied a graph-based approach using NetworkX (24) to construct a functional relationship network, enabling structural and topological analysis of interconnections among Actions, features, and publishers within the GitHub Marketplace ecosystem.

**RQ2: How do the provider characteristics relate to the evolution of functional relationships in this SECO?**

To investigate this, we analyzed the characteristics of GitHub Action publishers to understand their role in SECO evolution. Providers were classified into two categories: Independent Tool Providers, who exclusively release Actions with no functional overlap, and Dependent Tool Providers, who release at least one overlapping Action. We further identified Early Contributors, providers who initially (at $t_0$) released an Independent Action that later (at $t_1$) exhibited functional overlap, capturing their early and influential role in ecosystem innovation (25). By examining how provider characteristics relate to the functional evolution of their Actions, we aim to clarify the impact of Early Contributors and innovative providers in driving change and innovation within the SECO.

**RQ3: How can the evolution of the functional relationships in GitHub be explained using predefined SECO migration patterns?**

To explore evolutionary trends in the SECO, we analyzed the migratory behaviors of functionalities to assess the influence of pre-existing patterns on the evolution of GitHub Actions. Building on the nine behavioral patterns proposed by Sarro et al. (26) in the context of mobile app stores, we examined whether similar functional migrations occur within the GitHub Marketplace. We mapped the observed functional changes to these patterns and empirically assessed their impact on developer behavior, feature evolution, and ecosystem growth. This analysis aimed to determine whether migration dynamics in mobile ecosystems extend to software development marketplaces and how they shape innovation in GitHub Actions.

By answering these RQs, this paper contributes by:

- Identifying functional relations and their evolution in SECOs using a graph-based model.

- Characterizing early contributors and their influence on SECO evolution.

- Identifying migration patterns among GitHub actions as they evolve.

- Large-scale dataset construction and release, including 6,983 CI Actions developed by 3,869 providers.

We begin with a summary of related work in Section 2, followed by the data collection process in Section 3. Section 4 outlines our methodology, and Sections 5 and 6 present and discuss the results. We address threats to validity in Section 7 and conclude in Section 8.

## 2. Related Studies

In this section, we summarize the related studies relevant to research on SECOs, followed by studies on the GitHub Marketplace and GitHub Actions. This work builds upon the existing body of literature developed over the years (27; 28; 4; 16; 29; 30; 31; 32; 33; 34; 35; 36; 37; 38; 39; 40; 41; 42; 43; 44; 8; 45; 46; 47; 48; 49; 50; 51; 52; 53; 54; 55; 56; 57; 58; 59; 60; 39; 44; 61; 62; 63; 36; 37; 61; 62; 63; 64).

### 2.1. Software Ecosystems

SECOs play a crucial role in fostering creativity throughout the software development process (65; 66). Joshua et al. performed a systematic review of SECOs, highlighting their key features, benefits such as fostering co-innovation and reducing costs, and challenges including security and infrastructure management (19). Bosch (67) defines a SECO as a commercial ecosystem that includes software solutions and services designed to enable, support, and automate activities and transactions within both social and business environments. In recent years, analyzing SECOs to gain insights into various phases of the Software Development Life Cycle (SDLC) has become an area of growing interest among researchers (6). Malcher et al. (68) explored open innovation (OI) practices used to manage requirements in SECO, where multiple actors interact through various communication channels. They identified 10 OI practices and 14 communication channels, intending to make the requirements management process in SECO more informal, open, and collaborative.

### 2.2. GitHub Ecosystem and Marketplace

Saroar et al. (4) further contributed to the field of SECOs by establishing GitHub Marketplace as a SECO, identifying its components, and conceptualizing them within the broader definition of a SECO. Furthermore, they thoroughly analyzed the GitHub Marketplace to bridge the knowledge gap between state-of-the-art and state-of-practice. Their research highlighted that Action providers publish multiple Actions, along with various statistical analyses of different attributes within the GitHub Marketplace. This was the sole study we found pointing to the GitHub Marketplace as an SECO.

### 2.3. GitHub Actions for Workflow Automation

Even though there are limited studies targeting the GitHub Marketplace, various studies focus on GitHub Actions by studying the GitHub repositories that use Actions to automate their workflow (69; 70). Golzadeh et al. (17) conducted a nine-year empirical study to explore the evolution of Continuous Integration (CI) services within GitHub repositories, discovering that Actions have suppressed third-party tools, such as Travis. This was further followed from the line of study from the same research group to evaluate the use of Actions for different automated tasks (71; 72). Khatami et al. (73) later performed a study to identify 22 workflow smells in GitHub Actions by analyzing 10,012 commits and developing automated scripts for detection. Results highlight the need for better contextual understanding and communication with developers. Valenzuela et al. (74) examined GitHub Actions workflow maintenance in 200 projects, highlighting bug fixes and CI/CD improvements as key factors hinting at improved resource planning, best practices, and enhanced tool features. Since then, the field has increasingly focused on studying the automated repositories using these Actions (75; 76) and the impact of using Actions on CI/CD processes (77). Saroar and Nayebi (16) investigated developers' perceptions of GitHub Actions, focusing on reusability and workflow file creation efficiency, through a survey of 90 developers. They found that the developers opt to create new Actions due to various reasons (e.g., Bugs). Decan et al. (78) conducted a study on the outdatedness of project workflows utilizing GitHub Actions. Their findings revealed that, despite regular updates to Actions, most workflows rely on outdated versions that have been in use for over seven months.

### 2.4. Feature Extraction & AI for Software

To study the functional evolution of features in the marketplace, we first need to extract features from the tools in the SECO. There have been various methods to extract features from app descriptions. The feature extraction using n-grams and agglomerative clustering by Harman et al. (79) is one of the most popular methods. However, with the rise of generative AI and Large Language Models (LLMs) in software-related tasks (80; 81; 82; 83; 84), we opted to use LLMs for feature extraction. LLMs have been demonstrated to offer a scalable and explainable alternative to costly human evaluations. Zheng et al. (85) used strong LLMs as judges for open-ended questions, revealing that models like GPT-4 can effectively approximate human preferences with over 80% agreement.

### 2.5. Functional Evolution & Network Analysis in SECOs

To the best of our knowledge, our study is the first to explore the functional evolution of Actions in GitHub Marketplace. However, this is not the first time functional evolution has been studied in SECOs. Sarro et al. (26) performed a study to understand the migratory behavior of the features in a SECO over 33 weeks among pre-defined categories of the SECO. They established nine migratory behaviors and introduced a framework for analyzing app feature lifecycles using data from Samsung and BlackBerry app stores. They found that intransitive features exhibit unique behaviors and highlight correlations between price, rating, and popularity, offering valuable insights for developers. Seidl and Aßmann (86) presented a metamodel to capture and analyze the evolution of variability in SECOs, offering insights into the structure and changes within these ecosystems over time. Additionally, graphs have been used to study different relations in SECOs. The study by Kim et al. (87), which explored the content of mobile applications in

```
1    name: "Action's Name" ………… # Required/ Must be unique
2    author: "Action creator" ……………………………… #Optional
3    description: "Description of action" ………………………… #Required
4    input: ………………………………………………………… #Optional
5        input_ID: ………………………………………………… #Required
6            description: "Input description" ……… #Required
7            required: "True/False" ………………………… #Optional
8            default: "default value" …………………………… #Optional
9            deprecationMessage: "Warning message"…… #Optional
10   output: ………………………………………………………… #Optional
11       output_ID: ……………………………………………… #Required
12           description: "Output description" ………………… #Required
13           value: "Only for composite actions" ………… #Required
14   runs: …………………………………………………………… #Required
15       using: "runtime used to execute the code"
16   branding: ………………………………………………………… #Optional
```

**Figure 2:** A template of `action.yaml` file encompassing all potential fields and sub-fields. For this project, our emphasis was specifically on collecting data from the description fields.

the App Store using text-mining-based network analysis, is one such example. They visualized the associations among categories and applications, presenting both macro-level category networks and micro-level app networks.

## 3. Data collection

Developers can self-categorize their Actions into up to three of the 18 categories available on GitHub Marketplace (88; 15). These categories represent high-level processes. GitHub Actions were developed initially to automate Continuous Integration (CI) processes, making this category the oldest and most established within the ecosystem. The history of CI-related Actions dates back to 2015, predating the introduction of GitHub Marketplace itself. Given this longer evolutionary trajectory, CI Actions provide a rich dataset for analyzing functional evolution and competitive dynamics within the SECO. For this study, we scraped data from 6,983 Actions in the "Continuous Integration" category as of January 2024. Valid Actions on GitHub Marketplace require an `action.yaml` file in their root repository (15); therefore, we excluded any Actions missing this file or lacking a connection to their GitHub repository. Since our study focuses on the evolution of functional relationships among Actions, we retained only those present at both $t_0$ (no later than October 2023) and $t_1$ (January 2024), resulting in a final dataset of 5,006 Actions.

We developed an in-house scraper using `BeautifulSoup` (89) to collect data from GitHub Marketplace. For each Action, we extracted metadata, including the Action name, short and long descriptions, publisher name, number of stars, repository link, and other relevant attributes. We then navigated to each Action's repository, cloned it, and retrieved the contents of its `action.yaml` file. GitHub Marketplace restricts access to only the first 1,000 Actions per category in its default listing, as noted by Saroar et al.(4). To circumvent this limitation, we leveraged GitHub Marketplace's sorting algorithms along with a set of search terms to maximize coverage. Specifically, we used bi-grams extracted for each category from Saroar et al.(4), along with a complete set of English letters (upper and lowercase) and numbers (0 - 9) to generate search queries. By combining sorting strategies and search queries, we successfully retrieved 6,983 out of 7,248 Actions (96.34%) in the "Continuous Integration" category, ensuring comprehensive coverage for this study.

Each Action consists of one or more functionalities, which are described in natural language within the Action description and similarly structured in the `action.yml` file, along with its inputs and outputs. The structured format of `action.yml` makes it a more reliable source for feature identification, as it provides a standalone representation of an Action as seen by a developer. A template structure of the `action.yml` file is presented in Figure 2. Each `action.yml` file contains seven fields: `name`, `author`, `description`, `input`, `output`, `runs`, and `branding`. Among these, `name`, `description`, and `runs` are mandatory, while `author`, `input`, `output`, and `branding` are optional. The `description` field provides an overview of the Action's functionalities, whereas the `input` and `output` fields define the required inputs and possible outputs of the Action, respectively. Each of these fields also contains its own set of optional and mandatory subfields. Since the description field is mandatory and used to generate the description in the GitHub Marketplace, developers provide detailed explanations of their functionalities within this field. Therefore, we chose to extract all description fields for our analysis.

The data, along with the scripts used for mining GitHub Marketplace and Action repositories, as well as the subsequent analysis, are made publicly available in our GitHub repository (90).

## 4. Methodology

In this section, we outline the methodology employed in this study, organized according to the research question.

To study the evolution of Actions, we defined two time points in the version history of Actions: $t_0$ and $t_1$, where $t_0 < t_1$. The point $t_1$ is uniform across all Actions and represents the most recent update of the Action data as of January 2024 (the date of the last data extraction from GitHub Marketplace). In contrast, $t_0$ varies for each Action and marks its initial release, which can be any time between early 2015 to October 2024 (We selected Actions with a minimum time difference of 12 weeks between $t_0$ - $t_1$). By assigning a unique $t_0$ to each Action, we forgo the exploration of the birth of new Actions. However, this approach allows us to remain focused on our primary objective: studying the evolution of functional relationships among existing Actions within the ecosystem.

### 4.1. Extracting Features Overtime (RQ1)

We extracted all available description fields, including the overall `description` and those found in the input and output sections of the `action.yml` file, as the primary source for feature extraction. Subsequently, we identified relationships between these features using a network identification model.

<div style="border:1px solid #000;">

**Feature Extraction Prompt**

***Capacity and Role***: Requirement Analyst
***Insight***: We are looking to extract software features and functionalities from software descriptions. software features define the capabilities of the software.
***Statement***: Extract system functionalities from a given list of descriptions
***Personality***: accurate and precise and make sure the features are supported by the descriptions.
***Experiment***: output all features in a Python list. ONLY output the Python list.

here is the descriptions: `{descriptions}`

</div>

**Figure 3:** Prompt template created by modifying the CRISPE template and used to extract features from Action descriptions.

### 4.1.1. Mining Features from Action Descriptions

From the `yaml` file, we took several steps to extract Action's functionality:

**Using LLM to extract features:** With the rise of generative AI, particularly Large Language Models (LLMs), in supporting Requirements Engineering (RE) tasks (80; 81) and their capabilities as alternatives to human evaluations (85; 91; 92; 93), we opted for LLM-based feature extraction over traditional methods (79). We evaluated several frameworks on a subset of 100 GitHub Actions and found that the hybrid approach CRISPE (94; 95; 96) produced the most meaningful results. Figure 3 illustrates the prompt used to extract features from our dataset. We tested multiple LLMs, primarily those available through the `Groq` API (97), including MIXTRAL and LLAMA 3. To ensure consistency, we set the temperature to zero. Given the inherent risk of hallucination in LLM-generated outputs, we incorporated human verification to improve accuracy.

**Adjusting a subset of features using human annotation:** To assess and refine LLM-extracted features, the first author manually reviewed 610 features from 100 randomly selected Actions. They identified inaccuracies in 17.38% of the extracted features, adjusting their mappings to better align with the corresponding descriptions.

**Fine-tuning LLM using human annotations:** Fine-tuning LLMs and implementing Retrieval-Augmented Generation (RAG) have been shown to enhance output accuracy while significantly reducing model hallucinations in software engineering tasks (98; 99; 100). Based on the manually corrected features, we fine-tuned the model to improve extraction accuracy. We optimized feature extraction through few-shot prompting and RAG. We designed structured prompts incorporating human-verified feature mappings as exemplars, allowing the model to generalize better across new descriptions.

In our approach, we maintained a repository of manually corrected features and leveraged embedding-based retrieval to dynamically fetch the most relevant examples for each input. This retrieval mechanism ensured that the model was exposed to high-quality, domain-specific feature mappings without requiring parameter updates.

**Refining features:** We ran the fine-tuned model on the descriptions to refine the extracted features. In this process, we applied SelfCheckGPT (101), a self-consistency evaluation method that detects potential hallucinations in LLM outputs. This tool assesses Natural Language Inference (NLI), which measures whether a generated feature logically follows from the provided description. A high NLI error rate suggests inconsistencies or hallucinated content, whereas a low NLI rate indicates stronger alignment between the input and output. Using SelfCheckGPT, we found that the LLAMA3-8B-8192 model achieved the lowest NLI error rate (4.96%), meaning only 4.96% of the extracted features were flagged as inconsistent with their original descriptions. This validation confirmed our model selection as it demonstrated the best reliability in feature extraction.

**Identifying Unique features:** The use of different terminologies among developers may lead to the same feature being expressed in different ways. This inconsistency can complicate subsequent analysis and not reflect the true nature of the ecosystem. Hence, to remove potential duplicates and combine differently worded features, first we preprocessed the features (converting to lowercase, removing stopwords and special characters, and lemmatizing) and then vectorized them with the `bert-base-nli-mean-tokens` model (102) to convert to a vector embedding. Then, we used the `cosine_similarity` function of Scikit_learn library (103) to calculate the Cosine similarity score among features extracted at both $t_0$ and $t_1$ to identify unique features in the entire datasets and established a set of "Unique Features" (Figure 4). We used the recommended cosine similarity of 90% (104) to identify high similarity among these features at the two snapshots of the same action and between the Actions. Furthermore, we manually checked the features using crowd-sourcing and combined the similar features as identified by human annotators.

### 4.1.2. Identifying Network of Functional Relations

During the initial analysis of the features extracted from the marketplace, we identified that these tools (and subsequently their providers) are interconnected via the features they offer. We refer to these connections as functional relations and argue that these relations can be classified into four distinct types: *Independent*, *Subset*, *Intersect*, and *Identical* (**RQ1**). These relationships are formally defined as follows:

***Definition 1*** (*Feature Set*). A tool $A$ can be defined as a set of features $f_1, ..., f_n$ that describes its functionalities:

$$A = F_A = \{f_1, f_2, f_3, ..., f_n\}$$

In other words, $F_A$ is the *feature set* of tool $A$.

***Definition 2*** (*Independent Tools*). $A$ is an independent tool if for every feature $f$ in feature set $F_A$, feature $f$ does not belong to any other tool's feature set; in other words, no other tool in the ecosystem has features offered by this tool:

$$A \in A_{independent} \iff \forall f \in F_A | f \notin F_A^{'}$$

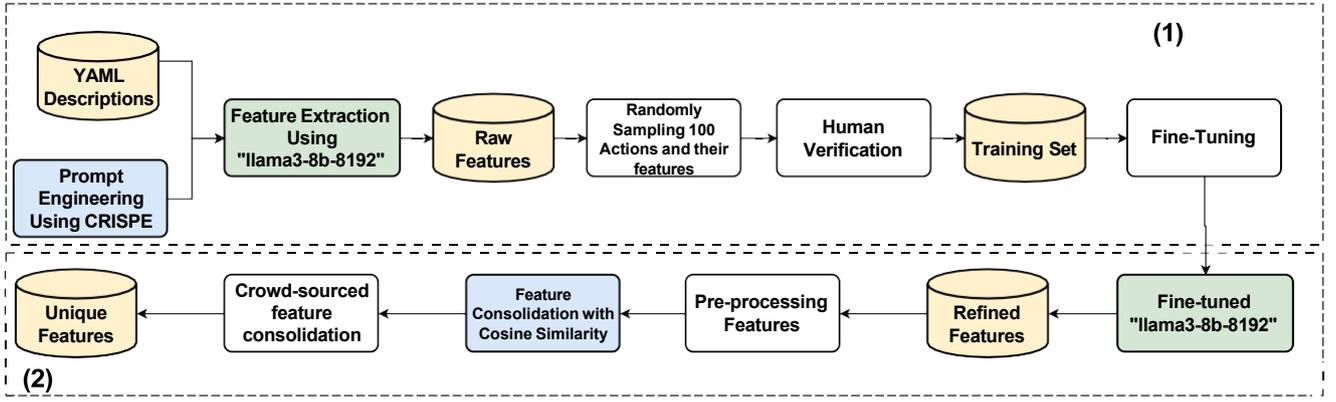

**Figure 4:** Feature extraction methodology using llama3-8b-8192 model fine-tuned using manually labeled sample dataset (1). Additionally, we used cosine similarity and crowd-sourcing to consolidate similar features (2).

where $A_{independent}$ is a set of all independent tools in the ecosystem.

**Definition 3** (*Subset Relation*). Consider tool $A$ with feature set $F_A$ and tool $B$ with feature set $F_B$. We define $A$ is subset of $B$ if all features in $F_A$ are also in $F_B$:

$$A \subset B \iff \forall f \in F_A | f \in F_B \land F_A \neq F_B$$

in other words, tool $B$ has all features of tool $A$ with some additional features.

**Definition 4** (*Identical Tools*). These types of tools are a special case of tools with a subset relation, where tool $A$ with feature set $F_A$ and tool $B$ with feature set $F_B$ have identical features:

$$A = B \iff F_A = F_B$$

**Definition 5** (*Intersect Relation*). For tool $A$ with feature set $F_A$ and tool $B$ with feature set $F_B$, we define $A$ intersects with tool $B$ if at least one feature in $F_A$ is also in $F_B$:

$$A \cap B \iff \exists f \in F_A | \begin{array}{l} f \in F_B \quad \land \\ F_A \neq F_B \quad \land \\ A \not\subset B \quad \land \\ B \not\subset A \end{array}$$

That is, tools $A$ and $B$ share at least one feature, but neither is a subset of the other, nor do they have identical feature sets.

We then employed the `NetworkX` and `PyVis` libraries to construct and visualize the network of the GitHub Marketplace ecosystem at two distinct time snapshots. `NetworkX` facilitated the creation and analysis of complex networks by allowing us to initialize directed graphs, add nodes representing individual Actions, and define directed edges to denote dependencies or interactions between them. Additionally, we applied various graph algorithms from `Networkx` to explore key structural properties, including `degree_centrality` (105) for identifying highly central nodes, `connected_components` (106) to uncover clusters of related actions, and to detect patterns of common functionalities across the network. Furthermore, the scalability of graph analysis tools allows for the efficient examination of large-scale Action networks for larger categories of the marketplace (e.g., CI category).

The most current network (at time $t_1$) is presented in Figure 5. The Actions, features, and publishers are depicted as nodes of green squares, blue triangles, and red circles, respectively, with edges connecting Actions sharing identical features and publishers. We studied this network to identify the relations and entities formalized in section 4.1.2 and section 4.2 and any other possible relations.

## 4.2. Understanding the role of Providers in the SECO (RQ2)

To examine how provider characteristics relate to the evolution of functional relationships within the GitHub Marketplace SECO, we conducted an empirical study using a longitudinal analysis (107; 27; 45) of publisher behaviors between two time snapshots, $t_0$ and $t_1$. Our methodology consists of three key steps: (i) identifying tool providers and categorizing them as independent or dependent, (ii) tracking their evolution over time, and (iii) analyzing the role of early contributors in shaping the ecosystem.

Using the established network in section 4.1.2, we identified two relationships between the providers and the Actions they release: *Independent Tool Provider* and *Dependent Tool Provider*. We formally define these as follows:

**Definition 6** (*Independent Tool Provider*). A provider in the SECO is an independent tool provider if and only if they publish independent tools. Consider provider $P$ publishing a set of tools $A_P = \{A_1, A_2, ..., A_n\}$, $P$ is Independent Provider if every tool in $A_P$ is an independent tool:

$$P \in P_{independent} \iff \forall A \in A_P | A \in A_{independent}$$

where $P_{independent}$ is a set of independent tool providers.

**Definition 7** (*Dependent Tool Provider*). A tool provider in the ecosystem is a dependent tool provider if it publishes at least one tool that is not independent. These providers publish tools that are either a subset of or intersect with other tools. Consider provider $P$ who has published a set of tools $A_P = \{A_1, A_2, ..., A_n\}$, $P$ is dependent provider if there is one tool in $A_P$ that is not an independent tool:

$$P \in P_{dependent} \iff \exists A \in A_P | A \notin A_{independent}$$

where $P_{dependent}$ is a set of dependent tool providers in the ecosystem.

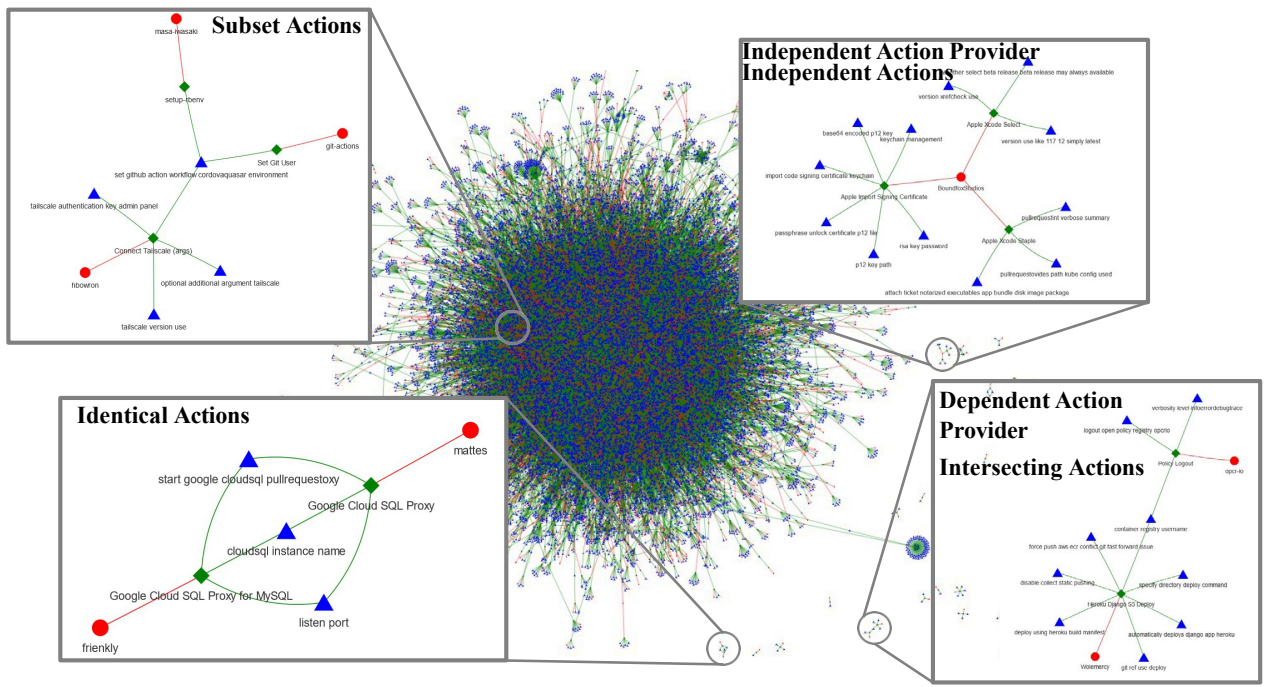

**Figure 5:** A highly connected network of Action functionality in the "Continuous Integration" category. The red circles represent each publisher connected to their released Actions (demonstrated in green squares), which are connected to their respective features in blue triangles. We identified the relations discussed in section 4.1.2 and section 4.2 and labeled them in the network.

To assess how provider relationships evolved, we tracked the functional status of each provider from $t_0$ to $t_1$. This analysis focused on how previously independent tool providers transitioned to dependent providers by incorporating or intersecting with existing functionalities. In the network model, the providers are represented as distinct nodes (depicted by red circles), each potentially surrounded by multiple Actions they have published.

### 4.3. Formalized Migratory Behavior in GitHub Marketplace (RQ3)

To address **RQ3**, we adapted the migratory behaviors outlined by Sarro et al. (26) to focus on inter-category migrations rather than the intra-category migrations used in their original study. They identified nine migratory behaviors for features across different categories at two time points, $t_0$ and $t_1$, separated by at least 33 weeks. We selected eight of these behaviors and redefined them to track migration across different Actions, as opposed to categories. This adjustment shifts the focus from inter-category to intra-category migration. We opted to use eight out of nine of these migratory behaviors since one of these behaviors (i.e."Unborn") is not in the context of our study. In the following section, we provide a summary of these eight behaviors and present a visualization in Figure 6.

**Weak Migration.** A feature has a weak migration if it was present in at least one tool at $t_0$ and migrated to at least one new tool at $t_1$. The feature may disappear from one or all tools it resided in at time $t_0$ (Figure 6 - a).

**Strong Migration.** Similar to weak migration, a feature has a Strong migration if it was present in at least one tool at $t_0$

and migrated to at least one new tool at $t_1$ with the difference that the features remain with all the tools it resided in at $t_0$ (Figure 6 - b).

**Weak Exodus.** This behavior is a special case of weak migration where a feature at $t_1$ disappears from at least one tool it resided in at $t_0$ and appears in at least one new tool at time $t_1$ (Figure 6 - c).

**Strong Exodus.** This behavior is another special case of weak migration where a feature at $t_1$ disappears from all tools it resided in at $t_0$ and appears in at least one new tool at time $t_1$ (Figure 6 - a). While all instances of strong exodus are cases of weak migration, the converse does not hold.

**Birth.** This behavior is a special case of Strong Exodus where the feature does not exist in any tools at $t_0$ and appears in one tool at $t_1$ (Figure 6 - d).

**Intransitive.** In this type of behavior, the feature at $t_1$ does not disappear from any tool it previously resided in, nor does it appear in any new tool.

**Weak Extinction.** A feature experiences weak extinction if it disappears from at least one tool at $t_1$ and does not appear in any new tool (Figure 6 - e).

**Strong Extinction.** The feature is said to have strong extinction if it does not appear in any new tool at $t_1$ and while disappearing from all tools (Figure 6 - f).

Sarro et al. (26) also defined undead or unborn features as a special case of strong extinction to be features that are outside of the ecosystem. However, we have not observed such relations in the GitHub marketplace. These defined migratory behaviors in GitHub Marketplace follow the same subsumption hierarchy as in the original work (26).

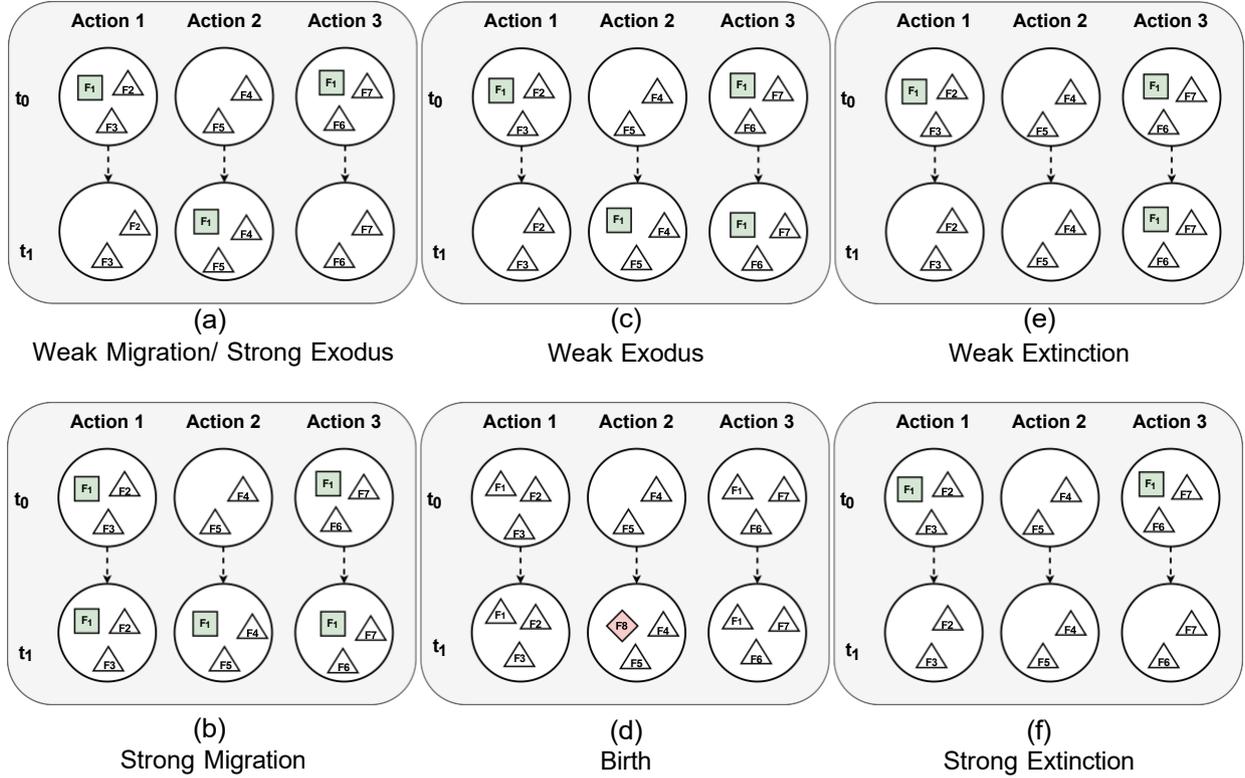

**Figure 6:** Modified Migratory behavior by Sarro et al. (26) where features migrate among tools **(RQ3)**.

## 5. Empirical Results

In this section, we analyze and discuss the results based on 10,694 features at time $t_0$ and 16,537 features at $t_1$, extracted from 5,006 Actions in the "Continuous Integration" category of the GitHub Marketplace (as described in Section 4), to address our three research questions.

### 5.1. Functional Evolution in GitHub Marketplace (RQ1)

To address this research question, we extracted functionalities from `yaml` descriptions, constructed a network using these functionalities, and identified relationships among Actions. In the following, we present our results for **RQ1**.

#### 5.1.1. Mining Functionalities from Action Descriptions

To analyze the functionalities of Actions, we extracted features from `yaml` descriptions using a combination of manual annotations and LLMs. We focused on extracting functional descriptions from `action.yml` files, as they provide structured information about each Action's capabilities.

Overall, we extracted 21,477 raw features at time $t_0$ and 38,583 raw features at time $t_1$, requiring further refinement. We refined the extracted features using a fine-tuned language model, reducing the feature set from 21,477 to 13,994 (34.84% reduction) at $t_0$ and from 38,583 to 20,569 (46.69% reduction) at $t_1$. Further manual refinements through crowd-sourcing resulted in 10,694 unique features at $t_0$ and 16,537 unique features at $t_1$, which were used in functional analysis.

#### 5.1.2. Identifying Network of Functional Relations

At $t_0$, we identified 10,694 unique features, with Actions containing between one and 104 features (an average of 3.42 features per Action). Additionally, we identified 1,186 disjoint components, where nodes are interconnected within a component but not connected to any nodes outside of it. Following the definitions provided in Section 4.1.2, we categorized Actions into 1,417 independent Actions, 503 subset Actions (i.e., a strict subset of another Action), 1,438 Actions that are identical to at least another Action, and 2,929 intersecting Actions (partially overlapping functionalities). Figure 7.a shows the distributions of these relations.

At $t_1$, the ecosystem had grown, containing 16,537 unique features compared to $t_0$, with each Action possessing between one and 78 features (mean: 4.93 features per Action). We identified 1,033 disjoint components, suggesting an increasingly interconnected ecosystem. The distribution of Actions evolved as 1,261 independent Actions, 364 subset Actions, 410 identical Actions, and 3,387 intersecting Actions, which shows a notable decrease in the number of subset and identical Actions, alongside an increase in intersecting Actions, suggesting a shift towards greater

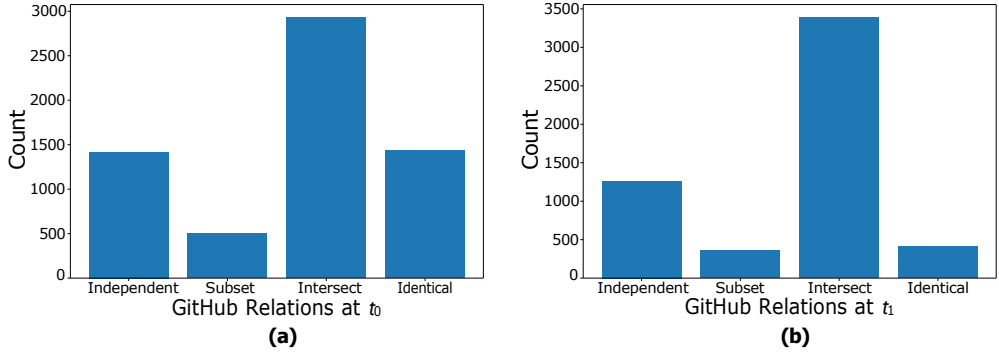

**Figure 7:** Count and distribution of relations at $t_0$ (a) and $t_1$ (b). Over time, the number of Actions that are subsets of other Actions and the number of Identical Actions have decreased, whereas the counts of Independent and Intersecting Actions have remained relatively stable with a slight increase (a).

feature integration and synergy among Actions. We depicted these distributions in Figure 7.b.

Over time, the GitHub Marketplace SECO has undergone significant structural changes, with increasing feature integration, reduction in redundancy, and growth in complexity. The overall number of features increased by 54.64%, while the number of disjoint functional components decreased by 12.90%, indicating a shift toward a more interconnected and cohesive ecosystem. Between $t_0$ and $t_1$, we observed an 11.01% reduction in independent Actions, alongside a 15.64% increase in intersecting Actions. This trend suggests that developers are recreating existing functionalities rather than creating entirely distinct Actions, contributing to a more interdependent SECO. Additionally, the average number of features per Action increased by 44.15%, reflecting growing functional complexity over time.

Interestingly, the number of subset Actions decreased by 27.63%, and identical Actions dropped sharply by 71.49%, indicating a gradual streamlining of redundant functionalities. This suggests a natural selection process, where functionally identical tools are either absorbed into broader offerings or evolve into more distinct, specialized tools.

Moreover, we found that Actions are not limited to a single relationship type; many exhibit multiple forms of interconnection, such as being both a subset and an intersecting Action simultaneously. These "co-occurring relations" contribute to a more intricate network, where Actions evolve dynamically within the SECO (Figure 5).

### 5.2. Providers' Role in Functional Relations (RQ2)

We define $t_0$ as the time at which each Action in our dataset is first introduced into the ecosystem. In total, we identified 3,867 publishers in the ecosystem at this time (first release of each Action). For this analysis, we ignore the birth and extinction of individual Actions and providers over time. In Section 4.2, we introduced two key provider-related terms: *Independent Tool Providers* and *Dependent Tool Providers*. Independent tool providers are those in the SECO who exclusively publish independent Actions, that is, all of their Actions are unique in terms of features and functionality. In contrast, dependent tool providers publish at least one Action that is not entirely unique but is interconnected with other Actions, either by being identical to, a subset of, or intersecting with them. At the time $t_0$, 1,043 of 3,867 publishers were classified as independent tool providers (26.97% independent and 73.03% dependent tool providers). However, by $t_1$, this number had decreased to 927, reflecting a 11.12% decline over time. This suggests that, over time, more publishers are relying on existing features within the ecosystem to create and maintain their Actions, with fewer publishers remaining completely independent in their offerings. However, instead of using the existing solutions that offer these features, they opt to recreate them. This trend may indicate a consolidation within the GitHub Marketplace, where the number of distinct tools diminishes and more interconnected tools emerge, hinting at growing competition and/or collaboration among providers. The slight increase in dependent tool providers from 2,824 at $t_0$ to 2,940 at $t_1$ further supports this notion of elevated competition and collaboration, as more publishers are replicating pre-existing features in the ecosystem.

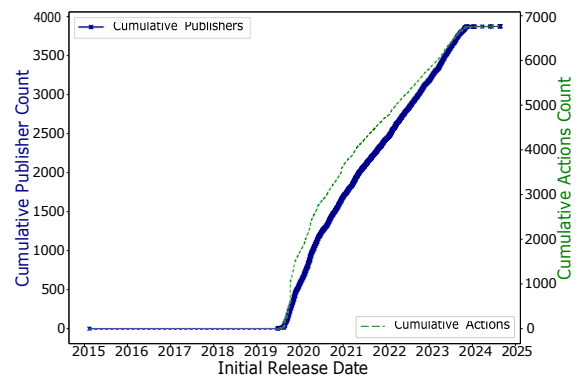

**Figure 8:** Number of publishers releasing the initial version of their Action over time. As time progressed, more new publishers and Actions joined the marketplace, whereas 85 publishers have been producing Actions ahead of their competition since 2015 **(RQ2)**.

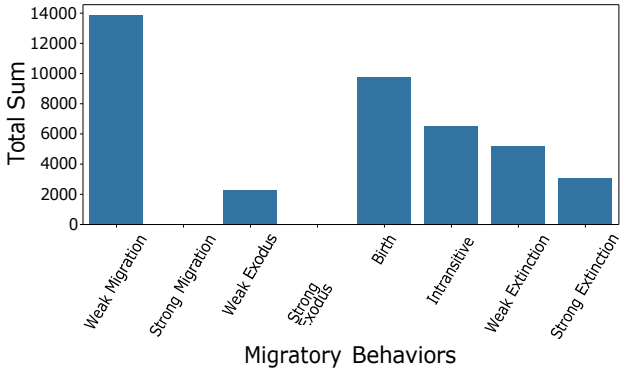

**Figure 9:** Total sum of features demonstrating each of the eight migratory behaviors **(RQ3)**.

### 5.2.1. Early Contributors in GitHub Marketplace

We found that even though the GitHub Marketplace was introduced in 2019, some publishers have been releasing Actions ahead of their competition from early 2015 to September 2019 (Figure 8). We therefore define these publishers who have released their Action before 2020 as "early contributors". We identified 85 early contributors in our ecosystem, releasing 90 actions between them. Among these, only four publishers release two or more actions before the rest of the competition. These publishers are "Azure", "garygrossgarten", "varunsridharan", and "warrenbuckley", none of which are subsets of any other Actions at $t_0$. Among these publishers, "varunsridharan" is the only one producing an independent Action while its other Actions intersect with 13 other Actions at $t_1$. On the other hand, the two Actions released by "Azur" intersect 11 times with different Actions, while 38 Actions have identical features to those in the two Actions released by "warrenbuckley". This indicates the popularity of the features offered by these Actions, resulting in developers reproducing the functionalities for their tools.

We identified eight independent tool providers among the early contributors (all releasing one Action, except one), out of which half remained independent at $t_1$, while the rest transformed into dependent tool providers. Furthermore, we identified 77 early contributors who are also dependent tool providers at $t_0$. The majority of these contributors release only one Action, with four contributors releasing two or more Actions. Among these contributors, seven transform into independent tool providers.

In summary, our analysis reveals a significant evolution in the ecosystem from $t_0$ to $t_1$. Initially, a substantial portion of publishers operated as independent tool providers; however, over time, the landscape shifted towards a greater reliance on interconnected features, as evidenced by the decline in independent offerings and the increase in dependent tool providers. This transition suggests a trend towards greater collaboration and competition within the GitHub

Marketplace, with publishers increasingly integrating and building upon existing features. The identification of "early contributors", who released Actions before the official marketplace introduction, underscores the role of early innovation in shaping the current ecosystem. Notably, while some early contributors remained independent, many others have transitioned towards a model of intersecting with other Actions to some degree, indicating that the Actions of these early contributors have become a building block for other Actions, with developers recreating the features offered by these early contributors. Thus, the results suggest that as the marketplace matures, the increasing interplay between independent and dependent tool offerings reflects a shift towards a more interconnected and dynamic ecosystem. This reshaping of the competitive and collaborative landscape aligns with the observed patterns of integration and adaptation among publishers.

### 5.3. Migratory Behavior of Actions in the SECO (RQ3)

To address RQ3, we identified a total of 19,099 unique features within the SECO. For each feature, we analyzed its presence at two time points, $t_0$ and $t_1$, to classify its migratory behavior according to the eight types defined in Section 4.3. Figure 9 illustrates the frequency of each migratory behavior observed among the features in the ecosystem. Our analysis shows that Weak Migration is the most common behavior, followed by Birth, Intransitive, Weak Extinction, Strong Extinction, and Weak Exodus. We did not observe any Strong Migration or Strong Exodus. This pattern indicates that migration and growth (Birth) are the dominant dynamics within the SECO. In contrast, intransitive behaviors and extinctions occur frequently. Still, they are not as dominant, suggesting that the SECO is in a growth phase with new features and products continually entering the market and existing ones being modified.

We also calculated the correlation between the functional relationships discovered in the SECO and the migratory behaviors (Figure 10). Our findings reveal a high correlation among migratory behaviors (likely due to their hierarchical relationships (26)). Additionally, there is a moderate correlation between an Action being independent at $t_0$ and remaining independent at $t_1$, as well as between an Action having intersections at $t_0$ and also having intersections at $t_1$. Actions with features exhibiting Weak Migration or Weak Exodus behaviors are more likely to evolve to intersect with other Actions and less likely to remain independent.

Furthermore, we analyzed the distribution of each migratory behavior across all Actions (Figure 11). Our findings show that every Action contains at least one feature exhibiting one of these migratory behaviors. Specifically, the behaviors associated with growth, such as Weak Migration and Birth, are the most dominant, with the highest average occurrences per Action. In contrast, the absence of Strong Migration and Strong Exodus indicates that features are highly dynamic, rarely remaining within or completely disappearing from their initial host.

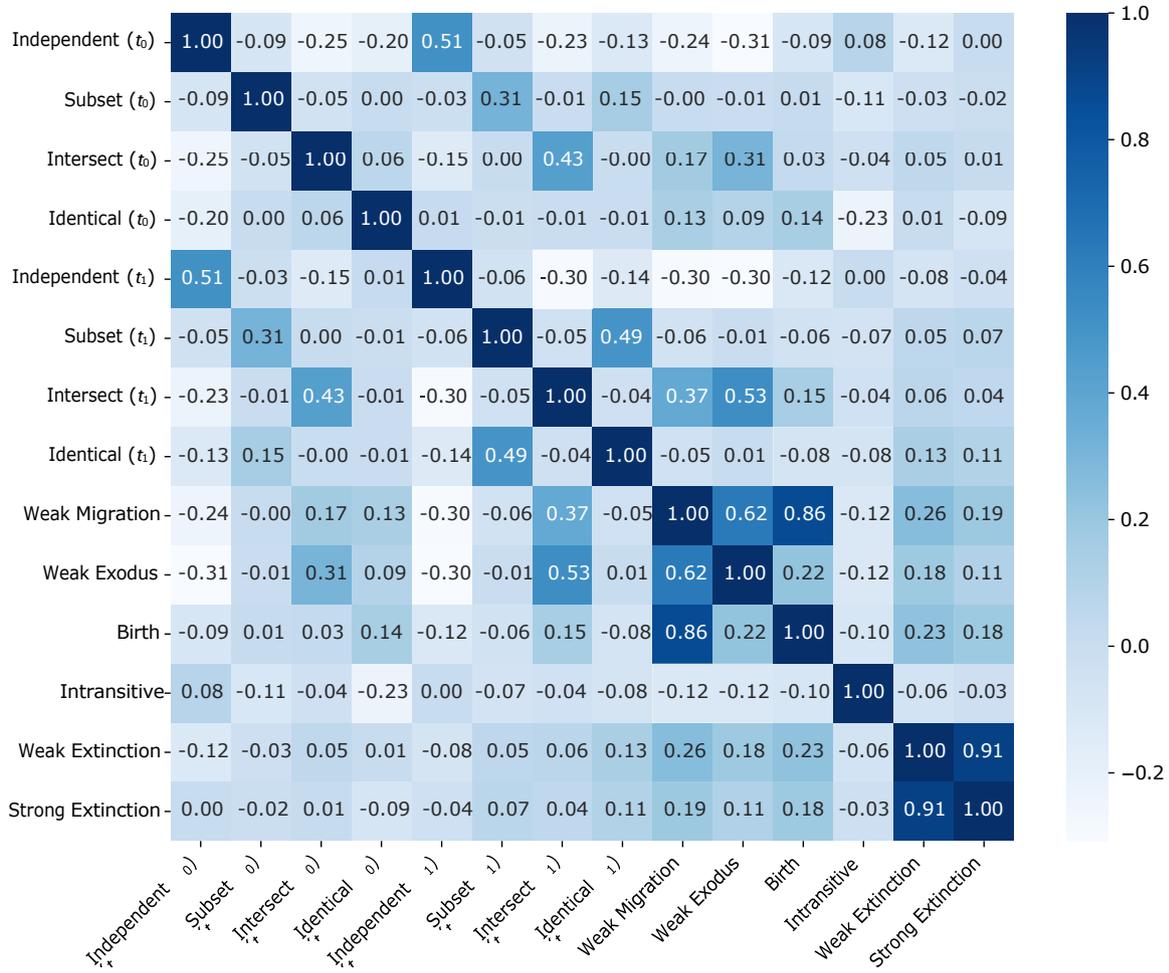

**Figure 10:** Correlation among functional relations and migratory behaviors in the SECO **(RQ3)**.

In summary, our analysis of the SECO reveals that migration and growth behaviors are dominant, reflecting an ecosystem that is actively evolving with new features and products entering the market. The dominance of Weak Migration and Birth behaviors demonstrates the dynamic nature of the ecosystem, where features frequently transition to new Actions and new features emerge regularly. Meanwhile, the absence of Strong Migration and Strong Exodus behaviors suggests that while features are highly mobile, they rarely remain with or entirely leave their initial host. The moderate correlations observed between functional relationships and migratory behaviors further indicate that Actions with certain characteristics are more likely to exhibit specific types of migration. These insights provide a deeper understanding of the evolutionary processes within the SECO and highlight areas for future research to explore how these dynamics might influence the long-term stability and growth of the ecosystem.

## 6. Discussion & Future Works

GitHub Marketplace is an example of SECOs for developers to share their automation solutions in the form of Actions. Since its release in 2019, the marketplace has experienced an exponential increase in the number of Actions and providers. Even though there have been many studies concerning Actions and their use in GitHub-hosted projects, very few studies have targeted the GitHub marketplace itself. We performed an evolutionary analysis of the Actions in the marketplace to provide Action developers with knowledge of how their product might evolve and how they can employ these evolutionary patterns to maintain a successful product.

### 6.1. Evolutionary Analysis of Actions

In this study, we focused on "Continuous Integration" Actions available on the GitHub Marketplace. We identified 5,006 Actions for our analysis and extracted a total of 19,099 features at two time points in the version history of the Actions. We noted that the Actions have become more complex over time, with a higher average number of features per Action. Additionally, the Actions have been shown to evolve to have higher synergy with less number of independent Actions. Furthermore, this synergy is rarely the case where the Actions are a subset of or identical to each other and happen when their features tend to overlap. We also found that independent Actions are more likely to retain their status than other types of Actions, even though the number of such Actions is declining. Independent Actions represent the most novel form of creativity in the SECO, where the features are unique and uncommon. Hence, a decrease in independent Actions can be a sign that the existing Actions contain most of the required features by the developers, and creating an optimal workflow for automation is a matter of finding the "right" combination of Actions. Furthermore, we found that there are Actions in the SECO that are used as the foundation for other Actions, where developers reproduce the features offered by these Actions in addition to other features (new or preexisting). Even though we did not consider feature content and characteristics in this study, future work could explore these aspects in detail to uncover how specific feature types, such as performance optimizations or integration functionalities, drive ecosystem dynamics and user adoption. This could complement our findings and provide a deeper understanding of the role of feature evolution in shaping software ecosystems.

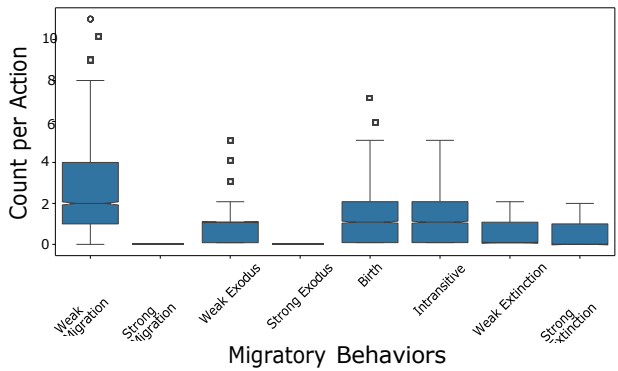

**Figure 11:** Distribution of migratory behavior in the Actions in SECO **(RQ3)**.

## 6.2. Super Action Concept

Considering the fact that the set of automation goals is finite and the existence of many Actions with overlapping features raises an interesting question regarding the value of developing a single comprehensive "Super Action" that automates multiple tasks simultaneously versus using independent Actions as modular building blocks to build highly customized workflows. In Figure 12, we illustrate a conceptual "Super Action", where the value of each feature (represented by bubble size) must strike a balance with the dependencies and synergies among co-occurring features. To better tailor and customize Actions for repository automation, it is crucial to develop methods for quantifying the synergy among tasks automated by GitHub Actions. Additionally, investigating the value and synergy inherent in these tasks is essential. Constructing the feature set for such Super Actions can be approached as a search problem, similar to the techniques used in release planning by Nayebi et al. (50). Answering these challenges and their implications is left for future studies, where one can explore whether creating a Super Action offers strategic advantages or if developers prefer to build their workflows using modular components like independent Actions. Overall, we found that the functional relations in the GitHub Marketplace SECO have evolved into a complex and intricate network where many Actions share common features.

## 6.3. Provider Patterns and early contributors

We extended our analysis by examining the influence of provider patterns on the evolution of functional relationships within the SECO. Our dataset revealed 3,869 providers, categorized into independent and dependent tool providers. Over time, the evolution of these functional relationships indicated a notable shift towards dependency, with more providers recreating existing features to maintain their products rather than creating novel, independent tools or using the existing solution. Additionally, we investigated the role of early contributors in driving this evolutionary process. We identified 85 early contributors, whose Actions exhibited significant overlap with other Actions over time, manifesting as subsets, intersecting, or identical relationships. These overlaps suggest that the feature sets developed by early contributors have become foundational for other providers, influencing the creation of new tools. The degree to which these feature sets are "good enough" or require further enhancement, as seen in SECO's evolution, remains an open question for future research.

## 6.4. Migratory Patterns of Features

We also explored the functional evolution of features through the lens of predefined migratory patterns as described by Sarro et al. (26). By analyzing 19,099 features at two points in the version history of Actions within SECO, we observed that features have evolved to exhibit behaviors consistent with growth and migration across Actions. Additionally, certain migratory behaviors, such as Weak Migration, Exodus, and Birth, were found to increase the likelihood of feature intersections among Actions. These behaviors suggest that features are not only evolving but also being redistributed or repurposed, reinforcing functional overlaps within the ecosystem. Future studies could delve deeper into the dynamics of these migratory behaviors by examining their long-term effects on tool innovation and diversity. Specifically, investigating whether features associated with Exodus or Weak Migration signal a decline in originality, or whether they act as stepping stones for functional synergy, could provide insights into how the SECO evolves to either foster or limit innovation.

In conclusion, our study provides a comprehensive evolutionary analysis of GitHub Marketplace's SECO, specifically focusing on "Continuous Integration" Actions. The findings highlight the increasing complexity of Actions, a growing reliance on reproducing existing features, and a shift towards dependent tools over time. Early contributors play a crucial role in shaping the ecosystem, with their feature sets serving as a foundation for future development. The migratory patterns of features further demonstrate how functionalities are repurposed, contributing to the ecosystem's dynamic nature. Future research should investigate the long-term impact of these patterns on innovation, particularly whether they drive or constrain the diversity of tools and features within the ecosystem.

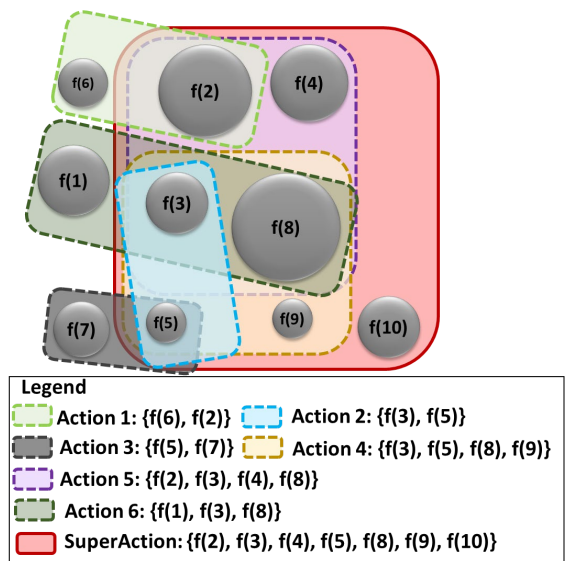

**Figure 12:** An example of the value of features and their co-occurrences and the concept of SuperActions.

## 7. Threats to Validity

In this section, we will examine threats to the validity of our study and results. During the study, we took precautions to ensure we were limiting the threats to the validity of results; however, our results are still under the influence of the following threats:

**Internal Validity** concerns whether the observed changes in our dataset over time can be attributed to genuine shifts in the ecosystem rather than confounding factors. We aimed to ensure internal validity by employing consistent methodologies for feature extraction and network analysis across both time points. The significant increase in the number of features and the decrease in disjoint components suggest substantial changes in the ecosystem's interconnectedness, supporting the validity of our findings.

Despite our efforts, there are potential threats to internal validity. We utilized Large Language Models (LLMs) for feature extraction, which, while leveraging state-of-the-art techniques to minimize hallucinations and enhance accuracy, may still introduce errors due to the inherent limitations of these models. Although we manually labeled a small sample to fine-tune the model and consolidated the extracted features using cosine similarity and crowd-sourcing, some inaccuracies may persist, potentially affecting our results. Additionally, the process of Action extraction involved using keywords, sorting algorithms, and web crawlers. Even though we managed to extract 96.34% of actions in the SECO, there is a possibility that some Actions were overlooked or not fully captured. This could lead to an incomplete dataset, which might influence the accuracy and completeness of our analysis. Thus, while our methodology strives to ensure robust internal validity, these potential issues should be acknowledged and considered when interpreting our findings.

**Construct Validity** ensures that our measures and analyses accurately capture the theoretical constructs we intend to study. In this analysis, we defined and categorized actions based on their feature relationships (e.g., independent, subset, intersecting) and tracked changes over time. Our use of network analysis to identify and categorize these relationships aligns with the theoretical constructs of feature interconnectedness and ecosystem complexity. However, the accuracy of these constructs relies on our theoretical definitions and the quality of our feature extraction. Future work could further validate these constructs by comparing them with other established metrics of ecosystem complexity and feature integration. Our reliance on the descriptions provided in the `action.yaml` file, which depends on developers accurately recording the functionalities of Actions, presents another risk. In some cases, developers may not have fully or correctly documented the functionalities, leading to discrepancies between the actual and recorded features of the Actions.

**External Validity** pertains to the generalizability of our findings beyond the specific dataset and time frame analyzed. Our study focuses on the GitHub Marketplace's "Continuous Integration" category, which may not fully capture the dynamics of other categories or comparable ecosystems. Although the observed trends, such as increased feature overlap and reduced redundancy, offer valuable insights into this specific domain, caution should be exercised when extending these conclusions to other categories or platforms. However, the identified relations and methodologies themselves can be applied to any SECO with minimal modification. To enhance the generalizability of our results, future research could explore whether similar patterns emerge across different categories or software ecosystems.

**Conclusion Validity** evaluates whether our statistical analyses support the conclusions drawn about the relationships and trends observed in the data. The observed increase in feature counts, changes in action relationships, and the evolution of independent and intersecting actions are based on rigorous statistical analyses of the dataset. The significant reductions in redundant and independent actions, along with the increase in feature overlaps, support our conclusions about the growing interconnectedness of the ecosystem. Nevertheless, ensuring the robustness of these conclusions requires consideration of potential limitations such as sample size, feature extraction, and the accuracy of our theoretical definitions. First, while our sample size is substantial, future research could benefit from larger and more diverse samples to enhance generalizability. Despite our efforts to maintain the consistency of extracted features, inaccuracies may still occur in feature identification and categorization. Additionally, the accuracy of our theoretical definitions is critical. Furthermore, we have assumed that the Action developers use the `action.yaml` as a technical interface and accurately record the functionalities offered by their actions in this document. However, any model limitations or assumptions might influence the observed trends. Addressing these factors through additional validation and refinement of our methods will be crucial for ensuring the robustness and reliability of our conclusions.

## 8. Conclusion

GitHub Marketplace is a unique source of data, as it provides access to both open-source data and developer activities. Utilizing this platform, we gathered 6,983 Actions from GitHub Marketplace to understand how the automation tasks are defined and connected among Actions in the SECO. Through our mining and graph analysis, we identified independent and overlapping Actions and found that Actions evolve to have more intricate interconnected functional relations. We identified 3,869 publishers and found that developers frequently utilize Actions created by early contributors as foundational elements for their products. Additionally, we analyzed 20,898 features and categorized them according to their migratory behavior. Our findings reveal high feature birth and migratory behavior (e.g., Weak Migration), indicating that features not only evolve but are also repurposed within Actions. This leads to an increasing interconnectedness within the SECO. In future studies, we aim to investigate the possibility of creating "Super Action" further by developing methods to measure the value and synergy among automated tasks within GitHub Actions, while evaluating the benefits of creating "Super Actions" versus customized workflows built from independent Actions.

## CRediT authorship contribution statement

**Elmira Onagh:** Conceptualization, Data curation, Formal analysis, Methodology, Visualization, Writing. **Maleknaz Nayebi:** Conceptualization, Funding acquisition, Supervision, Writing - review and editing.